\documentclass[aip,jcp,reprint,groupedaddress,a4paper]{revtex4-1}

\usepackage{amsmath}
\usepackage{txfonts}

\usepackage[T1]{fontenc}
\usepackage{textcomp}

\usepackage{bm}
\bmdefine{\bVector}{b}
\bmdefine{\BVector}{B}
\bmdefine{\eVector}{e}
\bmdefine{\EVector}{E}
\bmdefine{\fVector}{f}
\bmdefine{\FVector}{F}
\bmdefine{\gVector}{g}
\bmdefine{\pVector}{p}
\bmdefine{\PVector}{P}
\bmdefine{\qVector}{q}
\bmdefine{\QVector}{Q}
\bmdefine{\rVector}{r}
\bmdefine{\RVector}{R}
\bmdefine{\sVector}{s}
\bmdefine{\uVector}{u}
\bmdefine{\vVector}{v}
\bmdefine{\VVector}{V}
\bmdefine{\muVector}{\mu}
\bmdefine{\OmegaVector}{\Omega}

\usepackage[pdftex]{graphicx}

\usepackage{enumerate}

\draft 

\begin{document}


\title{Particle Monte Carlo simulation of string-like colloidal assembly in 2 dimensions} 



\author{Yuki Norizoe}

\author{Toshihiro Kawakatsu}
\affiliation{Department of Physics, Tohoku University, 980-8578 Sendai, Japan}


\date{20 November, 2011}

\begin{abstract}
We simulate structural phase behavior of polymer-grafted colloidal particles by molecular Monte Carlo technique. Interparticle potential, which has a finite repulsive square-step outside a rigid core of the colloid, was previously confirmed via numerical self-consistent field calculation. This model potential is purely repulsive. We simulate these model colloids in the canonical ensemble in 2 dimensions and find that these particles containing no interparticle attraction self-assemble and align in a string-like assembly, at low temperature and high density. This string-like colloidal assembly is related to percolation phenomena. Analyzing the cluster size distribution and the average string length, we build phase diagrams and discover that the average string length diverges around the region where the melting transition line and the percolation transition line cross. This result is similar to Ising spin systems, in which the percolation transition line and the order-disorder line meet at a critical point.
\end{abstract}


\maketitle 



%
%

%

\section{Introduction}
\label{sec:Introduction}
Colloidal particles are known as hard particles made from glasses, metals, and other materials. Latex is an example of such artificial colloids. Natural colloids, \textit{e.g.} pollen, are also seen in the environment. Colloids are known for their characteristic collective phenomena, for example glass transition and depletion interaction. Colloidal systems can be modelled by hard particle systems~\cite{Pusey}, owing to a solid and undeformable characteristic of the particles. The phase behavior of hard particle systems shows crystalization, which is called Alder transition~\cite{Alder1960,Alder1962}.

On the other hand, colloidal particles are liable to flocculate and make deposition after a long time, which disturbs colloidal dispersion. Polymers are, in industry, often grafted onto the colloidal particles to preclude the deposition and stabilize the dispersion. These are called polymer-grafted colloidal particles. We can control the interaction between polymer-grafted colloids by adjusting the length, species, the grafting density, and the chemical quality of the grafted polymers~\cite{Shull:1991,MatsenANDGardiner:2001,RoanANDKawakatsu:2002-1,BorukhovANDLeibler:2002,Akcora:2009}. Such usefulness of polymer-grafted colloids brings various applications and plays an important role in industry.

Simulating a simple model system via particle Monte Carlo simulation technique, we study the effect of the grafted polymers on the phase behavior of colloidal systems. In the present article, we show that particles interacting via spherically symmetrical repulsive square-step pair potential, which contains no attraction, self-assemble and finally align in strings. It is also shown that this string-like assembly indicates a similarity to percolation transition and critical phenomena~\cite{Norizoe:2005}.

A pair of polymer-grafted colloids interact with, in addition to their rigid cores, the grafted polymers outside these rigid cores. For example, let us consider an aqueous solution of the colloidal particles, diblock copolymers consisting of a long hydrophobic block and a short hydrophilic block are grafted. The hydrophobic blocks of these grafted diblock copolymers form a melt polymer brush, while the hydrophilic blocks form a thin shell swollen by the solvent. The rigid core is covered in the melt brush, which is covered in the swollen shell.

A pair of swollen brushes repels each other owing to the excluded volume effect~\cite{FleerAndOthers:1993}. On the other hand, slight attractive interaction potential is found between a pair of melt brushes upon contact~\cite{FleerAndOthers:1993}, since the polymer chain conformation relax when the two melt polymer brushes contact. The so-called Derjaguin approximation yields the effect of the curvature of the grafting surfaces on the interaction between grafted polymer brushes~\cite{Israelachvili}, provided that the radius of the curvature is far larger than the brush thickness. However, as long as the radius of the curvature is the same order of the brush thickness, no good analytical approximation is known. Numerical self-consistent field (SCF) calculation~\cite{Shull:1991,MatsenANDGardiner:2001,RoanANDKawakatsu:2002-1,FleerAndOthers:1993,BorukhovANDLeibler:2002} is necessary for such a situation.

Our SCF calculation has shown that, with some parameter sets, the interaction between a pair of our polymer-grafted colloidal particles reduces to a repulsive step potential containing no attraction~\cite{Norizoe:2005}. We approximate this potential by spherically symmetrical square-step potential with a rigid core,
\begin{alignat*}{2}
 & \phi (r) = \infty              & \qquad & r < \sigma_1,  \\
 & \phi (r) = \epsilon_0 \; (>0) &        & \sigma_1 < r < \sigma_2,  \\
 & \phi (r) = 0                   &        & \sigma_2 < r,
\end{alignat*}
where $r$ denotes the distance between the centers of the pair of the particles, $\sigma_1$ denotes the diameter of the colloids, and $\sigma_2$ is the diameter of the outer core formed by the polymer brush. Note that the height of the repulsive potential step $\epsilon_0$, which stems from the conformational entropy of the grafted polymers, is dependent on the temperature~\cite{Norizoe:2005}.

Due to the repulsive square-step in $\phi (r)$, the phase behavior of the present system depends on temperature. This is different from the behavior of the hard particle systems~\cite{Alder1960,Alder1962}. Boltzmann factor $\exp (-\epsilon_0 /k_B T)$ determines the probability that a pair of particles get on the step in the potential, where $k_B T$ denotes the thermal energy. The phase behavior at extremely low or high temperature is easily understood, although our model potential, $\phi (r)$, is invalidated in such extreme regions of the temperature.

At extremely high temperature, the phase behavior of our system is identical to the behavior of a system of hard spherical particles with diameter $\sigma_1$ since the height of the step, $\epsilon_0$, is vanishingly small compared with the thermal, or kinetic, energy~\cite{Norizoe:2005}.

At extremely low temperature, the repulsive step of $\phi (r)$ becomes far higher than the thermal energy. Due to this extremely high potential energy barrier, phase diagrams of the system is consistent with those of a system of hard spherical particles with diameter $\sigma_2$, when the system volume, $V$, is larger than the close-packed volume of the outer core, $V_0$. The pressure suddenly rises when $V$ is reduced across $V_0$, \textit{i.e.} the compression of the system box. This results in a phase transition at $V = V_0$, which is similar to the Alder transition~\cite{Norizoe:2005}.

In early works, a step potential with a rigid core or one with both a rigid core and a square-well were introduced as a model potential for the purpose of qualitatively simulating anomalies of phase behavior, \textit{e.g.} anomalous behavior of the melting curves and the critical behavior of phase diagrams, observed in systems of water, cesium, cerium, and others~\cite{Alder1977,Alder1979,DoubleStepPotential,Cesium1967,Cerium1970}.

At finite temperature $T$, the model system forms dimers, lamellae, and other various assembling structures~\cite{Malescio:2003,Malescio:2004,Norizoe:2003April,Norizoe:2003November,Master'sThesis,Norizoe:2005,Glaser:2007}. Glass transition~\cite{Fomin:2008} is observed in the same model system. These assemblies are also found in a system of particles interacting via continuous repulsive potential similar to our model potential $\phi (r)$~\cite{Camp:2003}. A variety of ground states of the same model system have been discovered at zero temperature via genetic algorithms~\cite{Pauschenwein:2008}. In the present study, we simulate our model system for the sake of constructing complete phase diagrams at finite $T$ in 2 dimensions. Simulation results in 3 dimensions will be published in our forthcoming article~\cite{NorizoeForthcomingPaperFrogNVT3D:2011}.

\section{Simulation methods}
\label{sec:SimulationMethods}
Monte Carlo simulations are performed in 2 dimensions in the canonical ensemble ($NVT$-ensemble) via the standard Metropolis algorithm~\cite{ComputerSimulationOfLiquids,Frenkel:UnderstandingMolecularSimulation2002}. $N$ and $V$ denote the number of particles and the system volume (area) respectively. Mersenne Twister is chosen as a random number generator for our simulations~\cite{MersenneTwister1,MersenneTwister2,MersenneTwister3}. In the initial state the particles are placed on triangular lattices in the system box with periodic boundary conditions. In one simulation step, a particle is picked at random and given a uniform random trial displacement within a square whose sides have length $0.4 \sigma_1$. A Monte Carlo step (MCS) is defined as $N$ trial moves, during which each particle is chosen for the trial displacement once on average. After $5.0 \times 10^6$ MCS, by which the system relaxes to the equilibrium state except at low temperature and high density, we acquire data every $10^4$ MCS and get 100 independent samples of particle configurations. Simulation results at $N=1200$ are mostly shown in the present article. We have confirmed, however, that the physical properties of the simulation system does not significantly change even when we run simulation for a larger system with $N=4800$.

$\sigma_1$ and $\epsilon_0$ are taken as the unit length and the unit energy respectively. $V_0$ ($V_0 = N {\sigma_2}^2 \sqrt3 /2$ in 2 dimensions) is used as the unit of the volume (area). Dimensionless temperature is defined as $k_B T/\epsilon_0$.

\section{Simulation results at $\sigma_2 / \sigma_1 = 2.0$}
\label{sec:SimulationResultsAtStepWidth2.0}
Here we discuss simulation results at $\sigma_2 / \sigma_1 = 2.0$, a typical case for the strongly curved grafted-brush.

\begin{figure*}[!tb]
\centering
\includegraphics[clip]{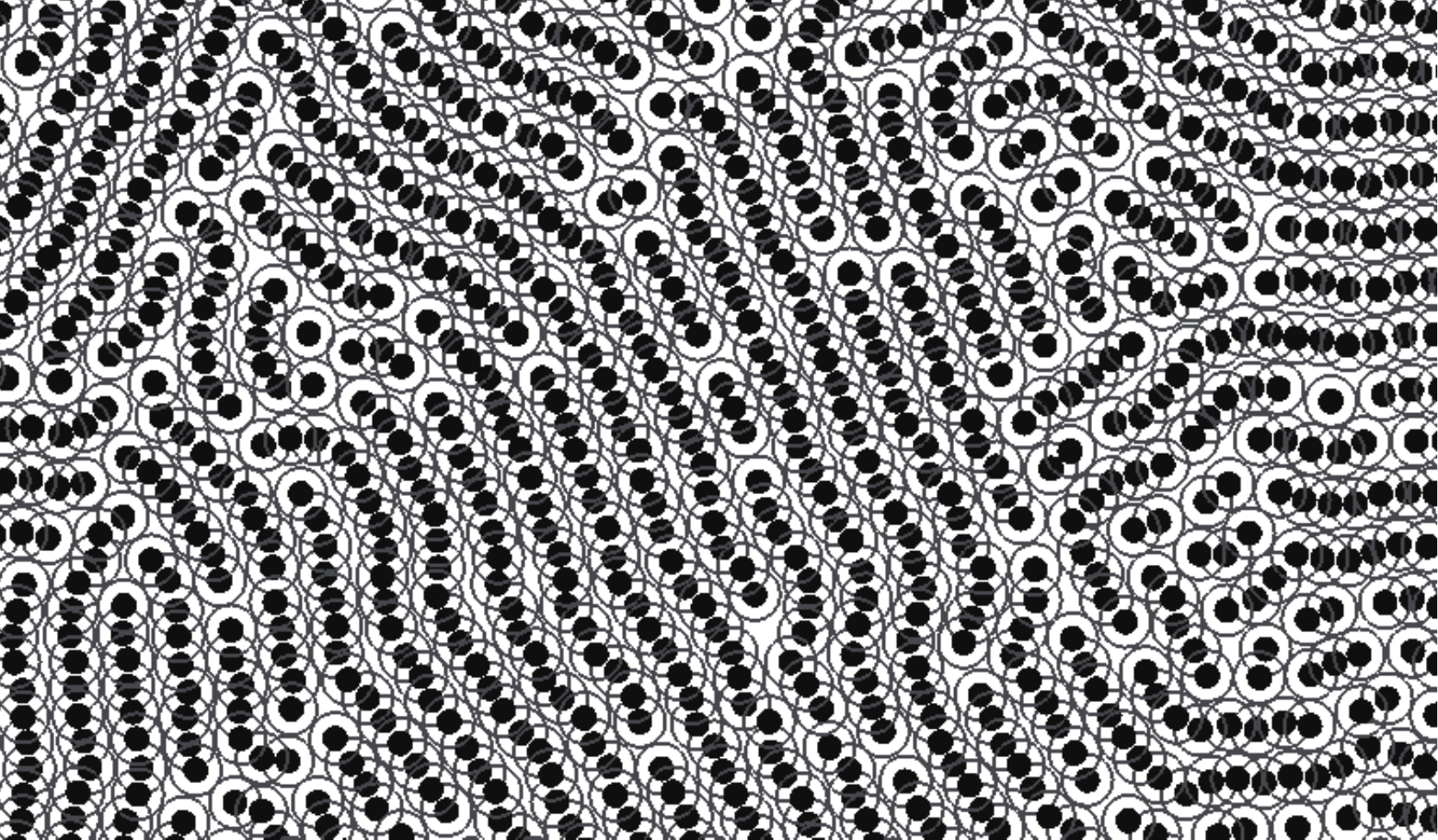}
\caption{Snapshot of a part of the system at $\sigma_2 / \sigma_1 = 2.0$, $N=10800$, $V/V_0=0.7$, $k_B T/\epsilon_0=0.1$, and $5.5 \times 10^{6}$ MCS.}
\label{fig:Frog2DNVTS20V07T01N10800_005500000MCS}
\end{figure*}

As a particular example, we show a snapshot of the system at $N=10800$, $V/V_0=0.7$, $k_B T/\epsilon_0=0.1$, and $5.5 \times 10^{6}$ MCS in Fig.~\ref{fig:Frog2DNVTS20V07T01N10800_005500000MCS}. We find that the colloidal particles self-assemble and finally align in strings. We called this effect ``frogspawn effect'' owing to the characteristic particle configuration and called this structure ``string-like assembly''~\cite{Norizoe:2003April,Norizoe:2003November,Master'sThesis,Norizoe:2005}. This effect has experimentally been confirmed lately~\cite{Osterman:2007}. This string-like assembly in 2 dimensions is also observed at $\sigma_2 / \sigma_1 = 2.0$~\cite{Malescio:2004} or $2.5$~\cite{Malescio:2003}, where the system is cooled from high $T$ to low $T$, and in a model system that consists of particles interacting via continuous repulsive potential similar to $\phi (r)$ in 2 dimensions~\cite{Camp:2003}. We have discovered that the string-like assembly also appears in our model system in 3 dimensions~\cite{Norizoe:2005}, which will separately be discussed in our forthcoming paper~\cite{NorizoeForthcomingPaperFrogNVT3D:2011}.

\subsection{Definitions of string length and cluster size}
\label{subsec:DefinitionOfStringAndCluster}
For the purpose of characterizing the structure of the particle aggregation, string length and cluster size are defined.

First, a ``bond'' is defined as an overlap between a pair of particles. A single network of bonds consisting of $M$ particles is defined as a ``cluster with the size $M$''. A particle that has more than two bonds in a cluster is defined as a ``junction'' of the network.

As is shown in Fig.~\ref{fig:Frog2DNVTS20V07T01N10800_005500000MCS}, there are clusters with the same size and the different configurations of the network. We introduce ``string length'' for quantitative distinction between the network configurations of clusters. A string is defined as a sequence of particles, each of which has exactly two bonds. The string terminates with junctions or particles with only one bond unless a circular configuration is composed. String length is defined as the number of particles along the string, including both the ends. A circular cluster consisting of three particles are excluded from the definition of the string although all the particles in such a cluster have two bonds.

\subsection{Stability of the string-like assembly}
\label{subsec:StabilityOfTheString-likeAssembly}
Here we check the dependence of the final colloidal assembly on the initial assembly. For this purpose, simulations with the same set of parameters as Fig.~\ref{fig:Frog2DNVTS20V07T01N10800_005500000MCS}, \textit{i.e.} $V / V_0 = 0.7$ and $k_B T / \epsilon_0 = 0.1$, have been performed starting from 4 different homogeneous and/or inhomogeneous initial configurations, which are perfect string-like assembly aligned in one direction, the close-packed configuration of the inner cores, and two other configurations.
%
The string length distributions and the cluster size distributions, which characterize the particle configuration found in the system, are analyzed for these 4 simulation runs with the same parameter set. The analyzed distributions indicate that all the simulation runs starting from different initial assembly result in the same string-like structure. This result demonstrates the stability of the string-like assembly. On the other hand, these cluster size distributions show an exponential distribution with the cluster size, which is similar to the formation of clusters in percolation systems in its non-critical region. This correspondence is discussed in section~\ref{subsec:PercolationPhenomenaAtStepWidth2.0}.

Considering the effect of the internal energy, we investigate the reason for the string-like assembly. At extremely low $T$, the particles tend to get down the step of the potential to decrease the internal energy. In regions of $V/V_0 < 1.0$, however, enough free volume to allow all the particles to get down the step is not found in the system. Therefore, at extremely low $T$ and $V/V_0 < 1.0$, the system builds long strings since a string-like linear cluster takes lower internal energy than a homogeneous crystalline structure, which is chosen as the initial configuration of our simulation. As the temperature increases, the string length decreases due to large probablity of overlaps between the particles at high $T$. Long strings are absent at $V/V_0 \approx 1.0$ since the particles seldom overlap. At high density, since the small free space rarely allows the particles to get down the step, the particles build only a few strings and most particles are junctions.

\subsection{Average string length at $\sigma_2 / \sigma_1 = 2.0$}
\label{subsec:AverageStringLengthAtStepWidth2.0}
\begin{figure}[!tb]
	\centering
	\includegraphics[clip]{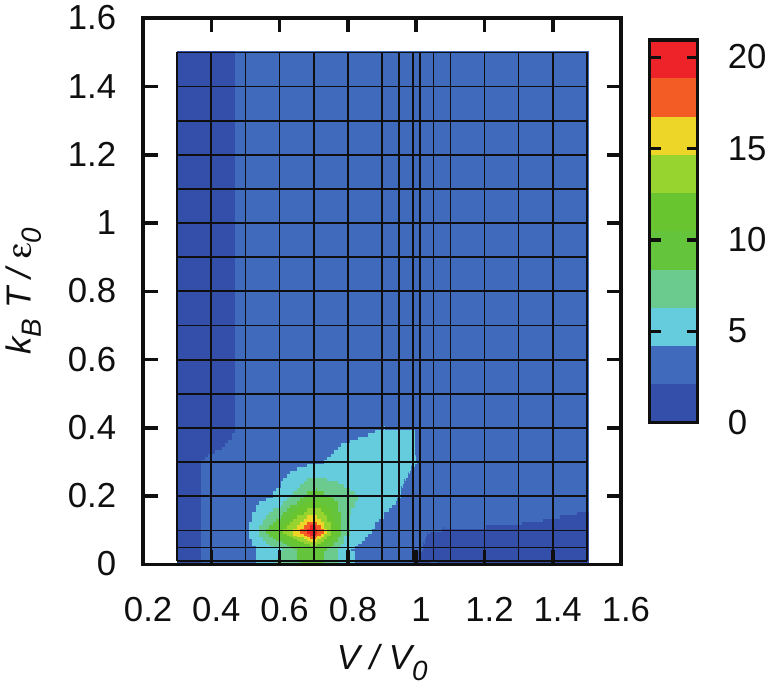}
	\caption{The average string length at $\sigma_2 / \sigma_1 = 2.0$ and $N=1200$. Blue regions represent short strings and red ones long strings. Grids denote the points where the simulation is performed.}
	\label{fig:Frog2DNVTS20-StringAverageLength}
\end{figure}

For the sake of locating the region of the string-like assembly, we analyze the average string length which is presented in Fig.~\ref{fig:Frog2DNVTS20-StringAverageLength}. The average string length has a sharp peak in a region of $0.1 < k_B T/\epsilon_0 < 0.2$ and $0.6 < V/V_0 < 0.7$. Long strings are observed only in the vicinity of this peak. At high density, there are only a few strings since particles are rarely allowed to get down the potential step from the initial crystalline configuration. At $V/V_0 > 1.0$ and low $k_B T/\epsilon_0$, particles seldom overlap and most particles are isolated.

The graph of the average string length with a resolution finer than Fig.~\ref{fig:Frog2DNVTS20-StringAverageLength} indicates that the maximum average string length exceeds 100 at $V/V_0 \cong 0.64$ and $k_B T/\epsilon_0 \cong 0.12$, which suggests a divergence of the string length~\cite{Norizoe:2005}. Simulation results via 3-reservoirs Monte Carlo method, which show parallel-arranged globally-anisotropic defect-free string-like assembly, corroborate this divergence~\cite{2010:NorizoeMuPTLetterArXiv,2011:NorizoeMuPTFullArXiv}. 3-reservoirs method allows us to efficiently simulate the equilibrium state. This result via 3-reservoirs method is discussed in section~\ref{subsec:ThermodynamicPhaseDiagramAtStepWidth2.0}.

\subsection{Percolation phenomena at $\sigma_2 / \sigma_1 = 2.0$}
\label{subsec:PercolationPhenomenaAtStepWidth2.0}
\begin{figure}[!tb]
  \centering
  \includegraphics[clip]{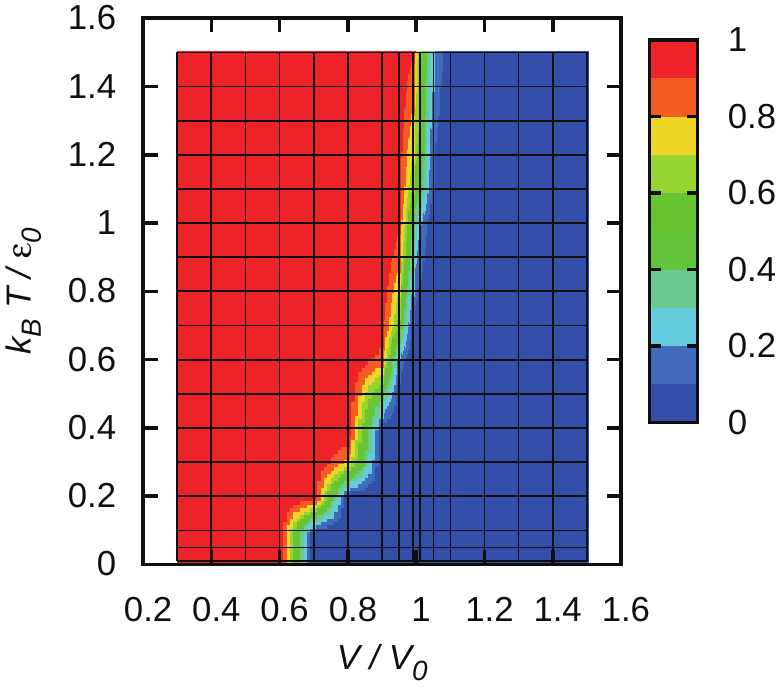}
  \caption{Occurrence probability of percolated clusters for the same system as Fig.~\ref{fig:Frog2DNVTS20-StringAverageLength}. $\sigma_2 / \sigma_1 = 2.0$ and $N=1200$.}
  \label{fig:Frog2DNVTS20-PercolationTotal}
\end{figure}

As is observed in Fig.~\ref{fig:Frog2DNVTS20-PercolationTotal}, at $N=1200$ and $k_B T/\epsilon_0=0.1$, the maximum cluster size suddenly changes between $V/V_0=0.6$ and $V/V_0=0.7$. At $V/V_0 = 0.7$ and $0.8$ the maximum cluster size found in the system equals no more than approximately 70 in our simulation runs, whereas at $V/V_0 = 0.6$ and $0.5$ exceeds 1000. At $V/V_0 \geq 0.7$ many small clusters are observed in the system, whereas at $V/V_0 \leq 0.6$ most particles in the system compose one large cluster. In order to understand this sudden change, we calculate the probability of the occurrence of a large cluster that bridges both the sides of the system box, i.e.~a percolated cluster~\cite{Stauffer1985}. The occurence probability of the percolated cluster at each $k_B T/\epsilon_0$ and $V/V_0$ is presented in Fig.~\ref{fig:Frog2DNVTS20-PercolationTotal}. A sharp boundary around $V/V_0 \approx 0.7$-$1.0$ divides percolation and non-percolation regions.
The occurrence probability of the percolated clusters changes abruptly at this sharp boundary, which suggests a similarity of this phenomenon to the usual percolation phenomena~\cite{Stauffer1985}.

At low density, $V/V_0 > 1.0$, the particles do not frequently overlap due to enough free volume to allow all particles to get down the interaction potential step. This results in the boundary in Fig.~\ref{fig:Frog2DNVTS20-PercolationTotal} asymptotically approaching to a line $V/V_0 = 1.0$ at high $k_B T/\epsilon_0$. On the other hand, at low $k_B T/\epsilon_0$, the boundary is located at $V/V_0 \approx 0.65$, between Alder transitions of outer cores and inner cores.

\begin{figure}[!tb]
	\centering
	\includegraphics[clip]{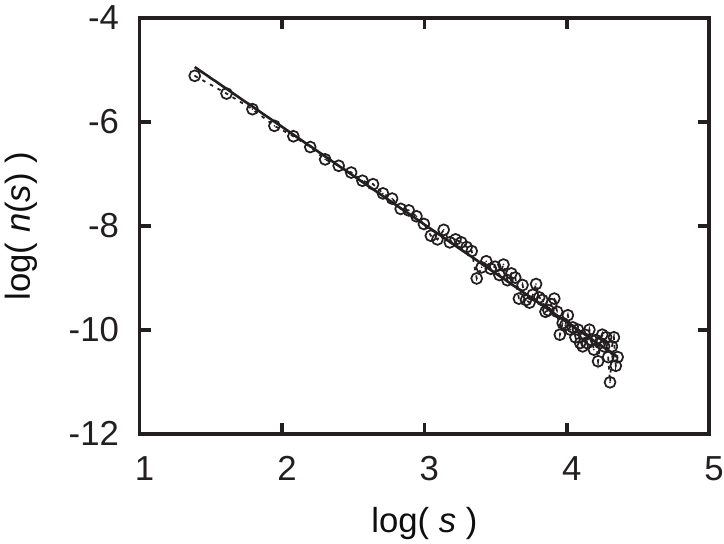}
	\caption{Cluster size distribution at $\sigma _2 / \sigma _1 = 2.0$, $V / V_0 = 0.98$, $k_B T / \epsilon_0 = 1.0$, and $N = 4800$. This parameter set is located on the percolation transition line in Fig.~\ref{fig:Frog2DNVTS20-PercolationTotal}. $\log (s)$-$\log (n(s))$ graph is plotted. A solid black line shows a linear fitting of this graph, $y = -\tau * x + b$ with $\tau = 1.9$ and $b = -2.3$.}
	\label{fig:Frog2DNVTS20V098T10N4800-ClusterSizeDistribution}
\end{figure}

\begin{figure}[!tb]
	\centering
	\includegraphics[clip]{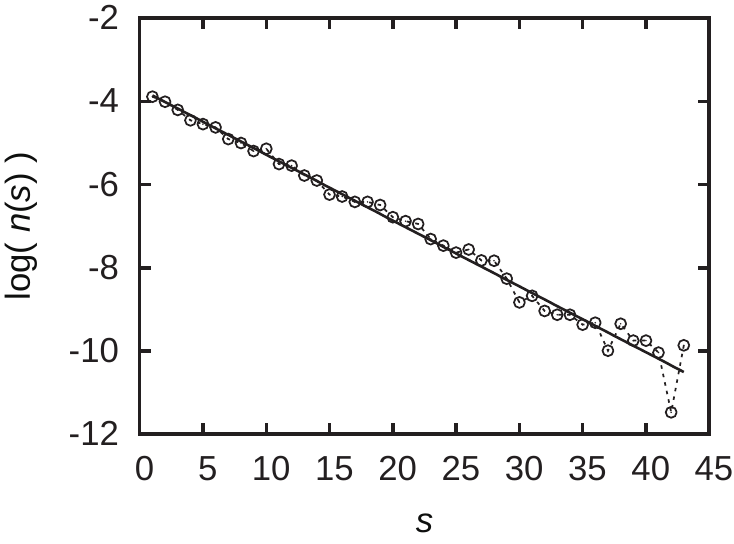}
	\caption{Cluster size distribution at $\sigma _2 / \sigma _1 = 2.0$, $V / V_0 = 0.7$, $k_B T / \epsilon_0 = 0.1$, and $N = 4800$. This parameter set is located outside the percolation transition line in Fig.~\ref{fig:Frog2DNVTS20-PercolationTotal}. $s$-$\log (n(s))$ graph is plotted. A solid black line shows a linear fitting of this graph, $y = -a * x + b$ with $a = 0.16$ and $b = -3.7$.}
	\label{fig:Frog2DNVTS20V07T01N4800-ClusterSizeDistribution}
\end{figure}

Here we study the relationship between our string-like assembly and the percolation phenomena. We calculate the cluster size distribution, $n(s)$, along the boundary curve in Fig.~\ref{fig:Frog2DNVTS20-PercolationTotal}. Here, the cluster size distribution is defined as, $n(s) = M(s) / N $,
where $s$ denotes the cluster size and $M(s)$ is the number of clusters with the size $s$ found in the system. An example of $n(s)$ on the percolation transition line in Fig.~\ref{fig:Frog2DNVTS20-PercolationTotal} is presented in Fig.~\ref{fig:Frog2DNVTS20V098T10N4800-ClusterSizeDistribution} in double logarithmic scales, which shows a relation, $n(s) \propto s^{-\tau}$ with $\tau = 1.9$,
where $\tau$ means the Fisher exponent, a critical exponent, of the percolation transition of our system. This power law with $\tau = 1.9$ is unchanged along the percolation transition line. Outside the percolation transition line, an exponential distribution is observed, which is also similar to the usual percolation phenomena. An example of this exponential distribution is plotted in Fig.~\ref{fig:Frog2DNVTS20V07T01N4800-ClusterSizeDistribution}.

Divergence of the string length in Fig.~\ref{fig:Frog2DNVTS20-StringAverageLength}, an anomalous phase behavior, is found around the percolation transition line. This result indicates a similarity between our string-like assembly and critical phenomena. One of the critical exponents, $\tau$, has been determined above in the present section, whereas, owing to the insufficient resolution of the data, other critical exponents are not evaluated in the present work.

\subsection{Thermodynamic phase diagram at $\sigma_2 / \sigma_1 = 2.0$}
\label{subsec:ThermodynamicPhaseDiagramAtStepWidth2.0}
A thermodynamic phase diagram is constructed in the present section.

\begin{figure}[!tb]
	\centering
	\includegraphics[clip]{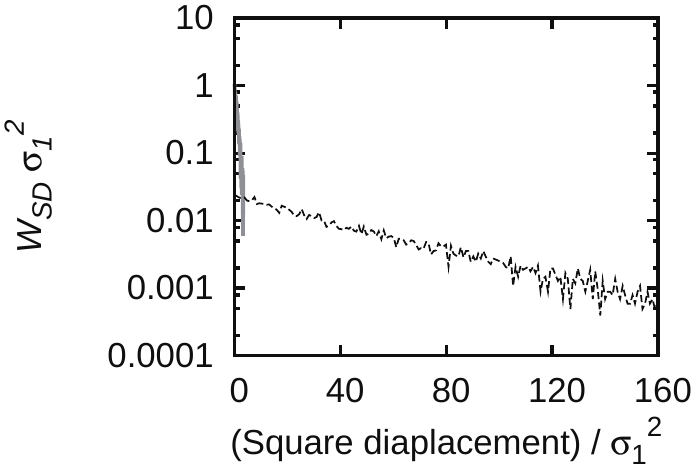}
	\caption{The probability density of the square diaplacement of the particles, $W_{\text{SD}}$, during $1 \times 10^5$ MCS at $\sigma _2 / \sigma _1 = 2.0$, $N=1200$, and $k_B T / \epsilon_0 = 0.01$. The result at $V / V_0 = 1.2$ is plotted by a grey solid line and $V / V_0 = 1.3$ by a black broken line. The number of analyzed particle configurations, \textit{i.e.} the number of samples, is equal to 9.}
	\label{fig:Frog2DNVTS20V12V13T001-MSDProbabilityDensityTotal}
\end{figure}

As has been discussed in section~\ref{sec:Introduction}, at extremely low temperature and $V / V_0 > 1.0$, the phase behavior of our system coincides with the behavior of the hard particle system of the outer cores. Alder transition of the hard particle system occurs at $V / V_0 \approx 1.3$ in 2-dimensions~\cite{Alder1962}. For the purpose of confirming this Alder transition of the outer cores of our colloids, in low temperature regions $k_B T / \epsilon_0 \le 0.1$, the probability density of the square displacement of the particles is analyzed at $V / V_0 = 1.2$ and 1.3. The results, \textit{e.g.} presented in Fig.~\ref{fig:Frog2DNVTS20V12V13T001-MSDProbabilityDensityTotal}, significantly change between $V / V_0 = 1.2$ and 1.3, which indicates the Alder transition. In other regions of $V / V_0$ and $k_B T / \epsilon_0$, a thermodynamic phase diagram is constructed based on the mean-square displacement of the particles (MSD) during $99 \times 10^4$ MCS. MSD at $V / V_0 = 1.3$ and $k_B T / \epsilon_0 = 0.01$ is chosen as the standard point, owing to the Alder transition of the outer cores. When MSD is smaller than this standard, the system is regarded as a solid. When MSD is larger than the standard value, the system is in a fluid phase or a coexisting phase of both the phases.

\begin{figure}[!tb]
	\centering
	\includegraphics[clip]{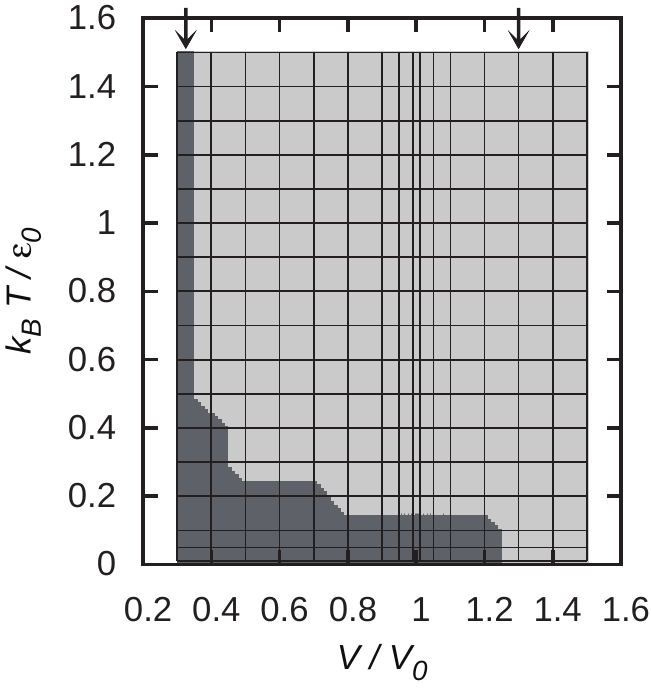}
	\caption{Thermodynamic phase diagram at $\sigma _2 / \sigma _1 = 2.0$ and $N = 1200$. Dark grey regions represent solid phases and light grey fluid or coexistence of both the phases. Alder transition points of the inner and the outer cores are pointed by arrows above the panel.}
	\label{fig:Frog2DNVTS20-ThermodynamicPhaseDiagram}
\end{figure}

The constructed thermodynamic phase diagram is given in Fig.~\ref{fig:Frog2DNVTS20-ThermodynamicPhaseDiagram}. At high $k_B T / \epsilon_0$ Alder transition of the inner cores of the colloids occurs in regions of $0.3 < V/V_0 < 0.4$, whereas Alder transition of the outer cores, \textit{i.e.} the transition at $V/V_0 \approx 1.3$, does not. At low $k_B T / \epsilon_0$ only Alder transition of outer cores occurs. At intermediate $k_B T / \epsilon_0$, the solid melts when not only $V/V_0$ but also $k_B T / \epsilon_0$ increases. These results are consistent with the phase behavior discussed in the introduction, section~\ref{sec:Introduction}.

The region where the string-like assembly, as is shown in Fig.~\ref{fig:Frog2DNVTS20V07T01N10800_005500000MCS}, and the percolation transition are located, \textit{i.e.} the region of $0.1 < k_B T/\epsilon_0 < 0.2$ and $0.6 < V/V_0 < 0.7$, is found in the solid phase between the two Alder transitions. This means that the string-like assembly is an amorphous solid, a non-equilibrium state.

\begin{figure}[!tb]
	\centering
	\includegraphics[clip]{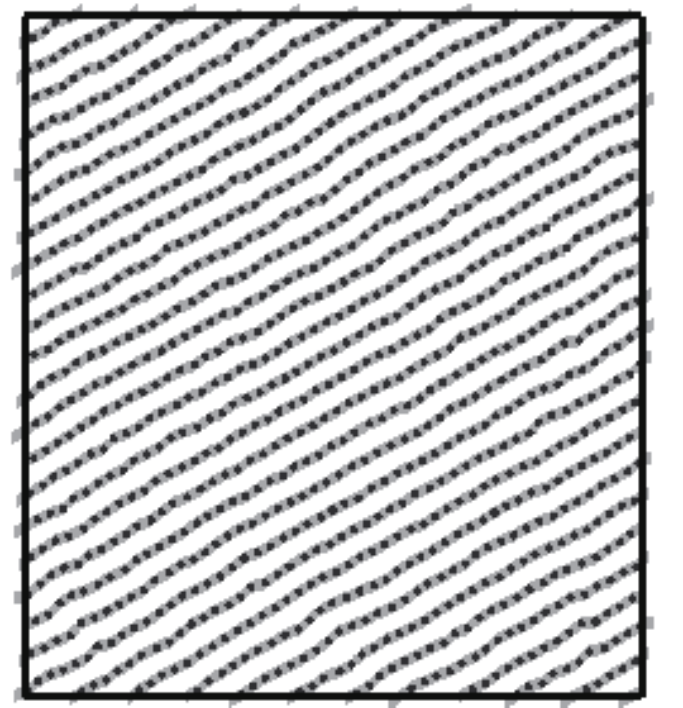}
	\caption{Snapshot of the system simulated via 3-reservoirs method~\cite{2010:NorizoeMuPTLetterArXiv}. Black dots represent the centers of the particles and grey lines denote the bonds between the particles.
	$k_B T / \epsilon_0 = 0.12 $, $P \sigma_1^{2} / \epsilon_0 = 1.1 $, and $\mu' = 29.93 $. 
	$P$ and $\mu'$ denote the pressure and the dimensionless chemical potential respectively.}
	\label{fig:Frog2DmuPTS20RD01e14P11T012_010000000MCS}
\end{figure}

Equilibration in the canonical ensemble demands a long compuational time. Therefore, the equilibrium state has recently been simulated via 3-reservoirs method, which provides shortcuts in the phase space from non-equilibrium to equilibrium states~\cite{2010:NorizoeMuPTLetterArXiv,2011:NorizoeMuPTFullArXiv}. Moreover, 3-reservoirs method allows the system itself to \textit{finely} and \textit{simultaneously} tune $N$ and the system box size to the equilibrated structure. Unlike other simulation methods, due to this unique feature, anisotropy of the equilibrated ordered structure is kept in a global scale. The periodicity of the structure is not constrained to the fixed system box size. Defects are removed from the ordered structure. The simulation results via this 3-reservoirs method have shown the globally-anisotropic defect-free string-like assembly as the equilibrated structure~\cite{2010:NorizoeMuPTLetterArXiv,2011:NorizoeMuPTFullArXiv}, presented in Fig.~\ref{fig:Frog2DmuPTS20RD01e14P11T012_010000000MCS}.

\section{Simulation results at various step widths}
\label{sec:SimulationResultsAtVariousStepWidths}
We have found the string-like assembly at $\sigma_2 / \sigma_1 = 2.0$. Here we study a range of $\sigma_2 / \sigma_1$, where the string-like assembly is observed.

When the step width $\sigma_2 / \sigma_1$ is taken to a limit, $\sigma_2 / \sigma_1 \to 1$, the potential step disappears and the phase behavior of the system is equivalent to the behavior of the hard particle system with a particle diameter $\sigma_1$. This means that the string-like assembly is absent in this limit. On the other hand, in a limit of $\sigma_2 / \sigma_1 \to \infty$, the inner hard core vanishes, which also results in the absence of the string-like assembly. Therefore, the string-like assembly is observed only in an intermediate region of step width $\sigma_2 / \sigma_1$. This result contradicts a conjecture by Malescio and Pellicane~\cite{Malescio:2004} that this assembly should be found at any step width satisfying a condition $\sigma_2 / \sigma_1 \approx 1$.

We simulate the system at various step widths and determine this intermediate region of $\sigma_2 / \sigma_1$ in the present section. The simulation results in section~\ref{sec:SimulationResultsAtStepWidth2.0} indicate this intermediate region ranges around $\sigma_2 / \sigma_1 = 2.0$, which simulation results in the present section corroborate. Other physical properties of the system at various step widths are also studied.

\subsection{Average string length at various step widths}
\label{subsec:AverageStringLengthAtVariousStepWidths}
\begin{figure*}[!p]
	\centering
	\includegraphics[clip]{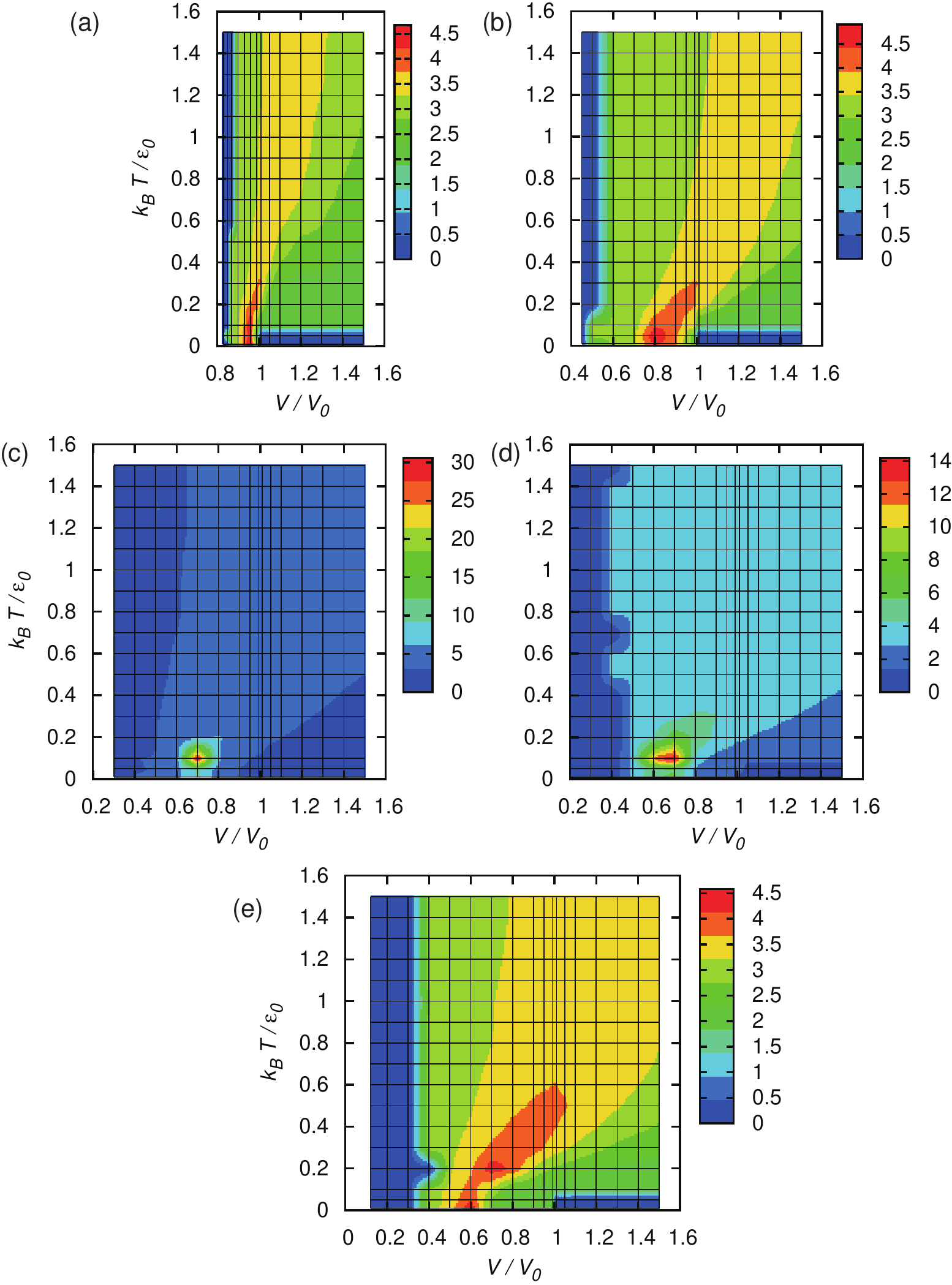}
	\caption{The average string length at $N=1200$. $\sigma_2 / \sigma_1 =$ (a): 1.1, (b): 1.5, (c): 1.9, (d): 2.5, and (e): 3.0. Red regions represent long strings and blue short strings. The average string length at $\sigma_2 / \sigma_1 = 2.0$ is given in Fig.~\ref{fig:Frog2DNVTS20-StringAverageLength}.}
	\label{fig:Frog2DNVTS11S15S19S25S30-StringAverageLength}
\end{figure*}

\begin{figure}[!tb]
	\centering
	\includegraphics[clip]{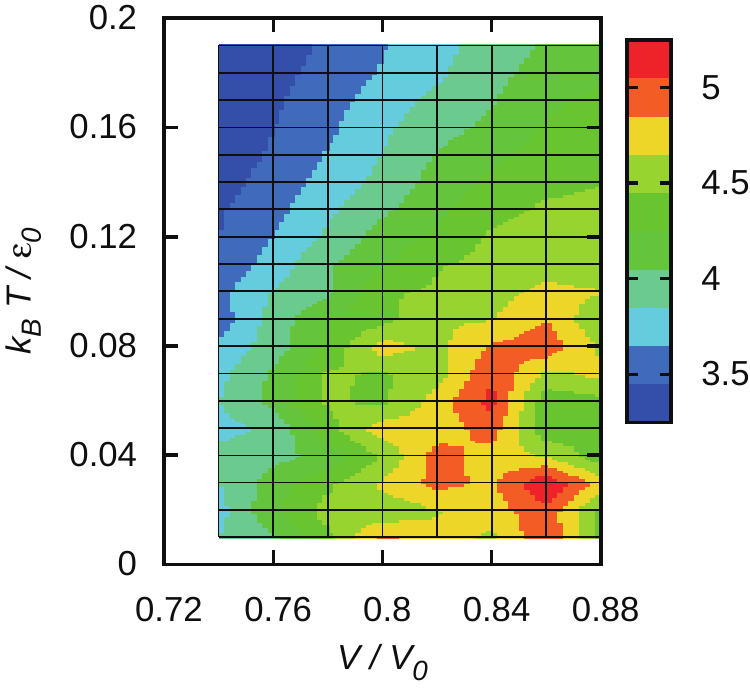}
	\caption{The average string length at $N = 1200$ and $\sigma_2 / \sigma_1 = 1.5$. The same graph as Fig.~\ref{fig:Frog2DNVTS11S15S19S25S30-StringAverageLength}(b), but with high resolution. The long strings and the divergence of the string length are absent at this $\sigma_2 / \sigma_1$.}
	\label{fig:Frog2DNVTS15Fine-StringAverageLength}
\end{figure}

\begin{figure}[!tb]
\centering
\includegraphics[clip]{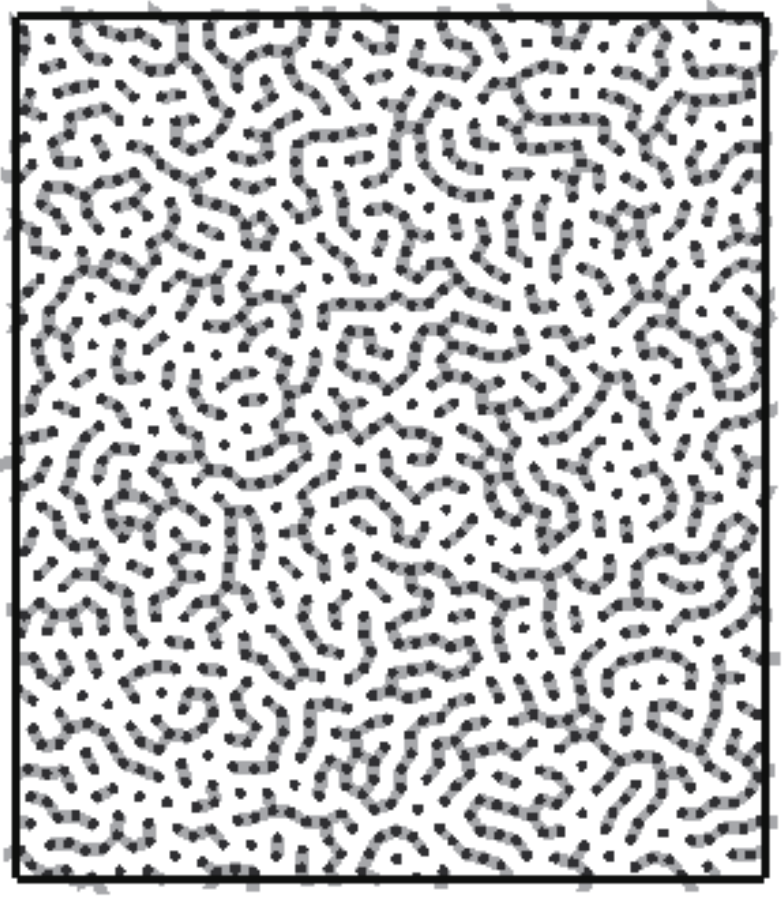}
\caption{Snapshot of the system on a peak of Fig.~\ref{fig:Frog2DNVTS15Fine-StringAverageLength}, \textit{i.e.} at $\sigma_2 / \sigma_1 = 1.5$, $N = 1200$, $V / V_0 = 0.84$, $k_B T/\epsilon_0 = 0.06$, and $5.0 \times 10^{6}$ MCS. Black dots represent the centers of the particles and grey lines denote the bonds between the particles. Long strings, as is shown in Fig.~\ref{fig:Frog2DNVTS20V07T01N10800_005500000MCS}, are not observed.}
\label{fig:Frog2DNVTS15FineV084T006_005000000MCS}
\end{figure}

The average string length at $\sigma_2 / \sigma_1 = 1.1$, 1.5, 1.9, 2.5, and 3.0 are given in Fig.~\ref{fig:Frog2DNVTS11S15S19S25S30-StringAverageLength}. At $\sigma_2 / \sigma_1 = 1.9$ and 2.5, the string-like assembly as is shown in Fig.~\ref{fig:Frog2DNVTS20V07T01N10800_005500000MCS} is observed, whereas, at $\sigma_2 / \sigma_1 = 1.1$, 1.5, and 3.0, the long strings are not observed in our simulation. The same graph as Fig.~\ref{fig:Frog2DNVTS11S15S19S25S30-StringAverageLength}(b), but with high resolution, is presented in Fig.~\ref{fig:Frog2DNVTS15Fine-StringAverageLength}, which demonstrates that the resolution of the graphs does not affect this result and that the divergence of the string length is absent at this $\sigma_2 / \sigma_1$. As an example, a snapshot of the system on a peak of Fig.~\ref{fig:Frog2DNVTS15Fine-StringAverageLength} is given in Fig.~\ref{fig:Frog2DNVTS15FineV084T006_005000000MCS}. Our simulation results indicate the region of the string-like assembly ranging in an interval $1.7 \lessapprox \sigma_2 / \sigma_1 \lessapprox 2.8$, \textit{i.e.} in the vicinity of $\sigma_2 / \sigma_1 = 2.0$.

\begin{figure}[!tb]
	\centering
	\includegraphics[clip]{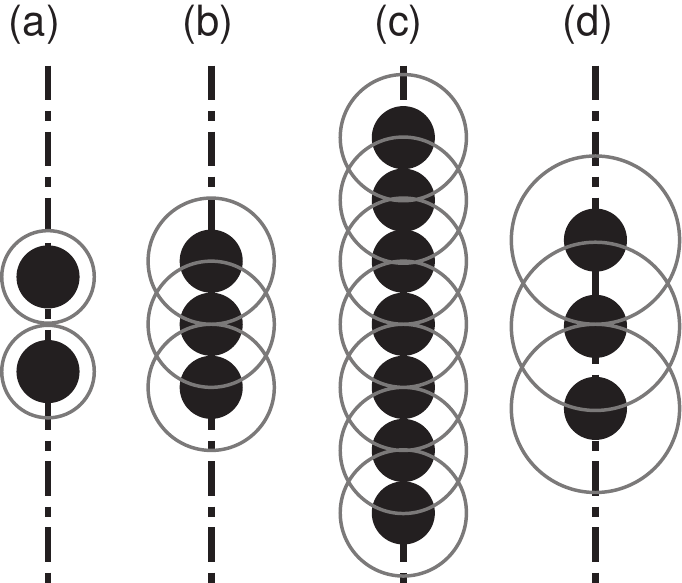}
	\caption{Sketch of local configurations of a few particles arranged on a straight line. (a): $\sigma_2 / \sigma_1 < 2.0$. (b) and (c): $\sigma_2 / \sigma_1 = 2.0$. (d): $\sigma_2 / \sigma_1 > 2.0$.}
	\label{fig:SketchStringsVariousStepWidths}
\end{figure}

The boundary of the step width, $\sigma_2 / \sigma_1 = 2.0$, divides characteristics of the local particle configuration. Below this boundary, \textit{i.e.} in the region of $\sigma_2 / \sigma_1 < 2.0$, the phase behavior of the system is similar to the phase behavior at the limit of $\sigma_2 / \sigma_1 \to 1$. When the outer cores of a pair of the particles are in close contact as is sketched in Fig.~\ref{fig:SketchStringsVariousStepWidths}(a), the 3rd particle is disallowed to be placed on a straight line between the pair, due to overlaps of the inner cores.

On the other hand, at this boundary of the step width $\sigma_2 / \sigma_1 = 2.0$, the 3rd particle is allowed to be arranged on this straight line, as is illustrated in Fig.~\ref{fig:SketchStringsVariousStepWidths}(b). In addition, in this particle configuration, the inner cores of the 3 particles are in close contact. The displacement of the 3rd particle from this position is limited to the direction normal to this straight line. The 3rd particle is pinched and confined by the other 2 particles. When particles are added to these 3 particles and closely packed along the straight line (Fig.~\ref{fig:SketchStringsVariousStepWidths}(c)), not only the inner cores but also the outer cores of all the particles, excluding both the end particles, simultaneously pinch and confine each other in the same way. This means that the inner core is pinched by the nearest neighbors and that the outer core by the next nearest neighbors. Therefore, the string-like assembly is stable in the direction of the straight line. When these straight strings are closely arranged in the same direction, the string-like assembly is also stable in the lateral direction since an additional energy cost is required to move and force one particle of a string into the adjacent string. Therefore, the system tends to keep the linear, \textit{i.e.} string-like, assembly at $\sigma_2 / \sigma_1 = 2.0$.

For the case, $\sigma_2 / \sigma_1 > 2.0$, the 3rd particle is able to be arranged on the straight line between the pair of the particles (Fig.~\ref{fig:SketchStringsVariousStepWidths}(d)). However, due to the free space between the pair, the 3rd particle is allowed to freely move. This disrupts the string-like assembly. When the inner cores of the 3 particles are closely packed along the straight line, this particle configuration demands $\epsilon_0$ additional energy cost, which also disrupts the string-like assembly.

In 3 dimensions, additional degrees of freedom reduce the pinch and the confinement of the particles and decrease the interval of $\sigma_2 / \sigma_1$ where the string-like assembly is observed~\cite{NorizoeForthcomingPaperFrogNVT3D:2011}. In other words, in 3 dimensions, the string-like assembly is observed only in the close vicinity of $\sigma_2 / \sigma_1 = 2.0$.

\subsection{Percolation phenomena and thermodynamic phase diagrams at various step widths}
\label{subsec:PercolationPhenomenaAndThermodynamicPhaseDiagramsAtVariousStepWidths}
\begin{figure*}[!p]
	\centering
	\includegraphics[clip]{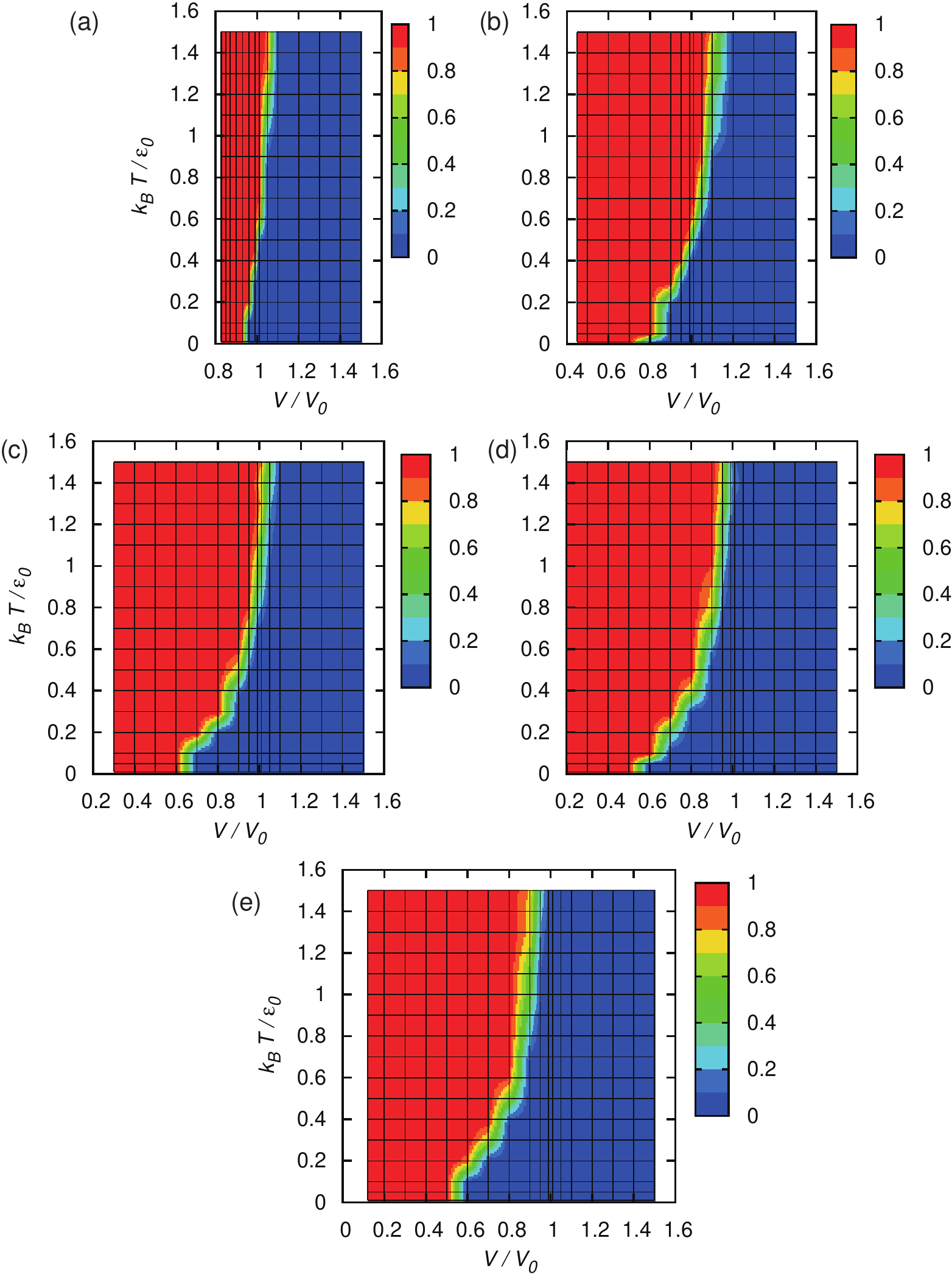}
	\caption{Occurrence probability of percolated clusters at $N=1200$. $\sigma_2 / \sigma_1 =$ (a): 1.1, (b): 1.5, (c): 1.9, (d): 2.5, and (e): 3.0. Corresponding occurrence probability of percolated clusters at $\sigma_2 / \sigma_1 = 2.0$ is given in Fig.~\ref{fig:Frog2DNVTS20-PercolationTotal}.}
	\label{fig:Frog2DNVTS11S15S19S25S30-PercolationTotal}
\end{figure*}

\begin{figure*}[!p]
	\centering
	\includegraphics[clip]{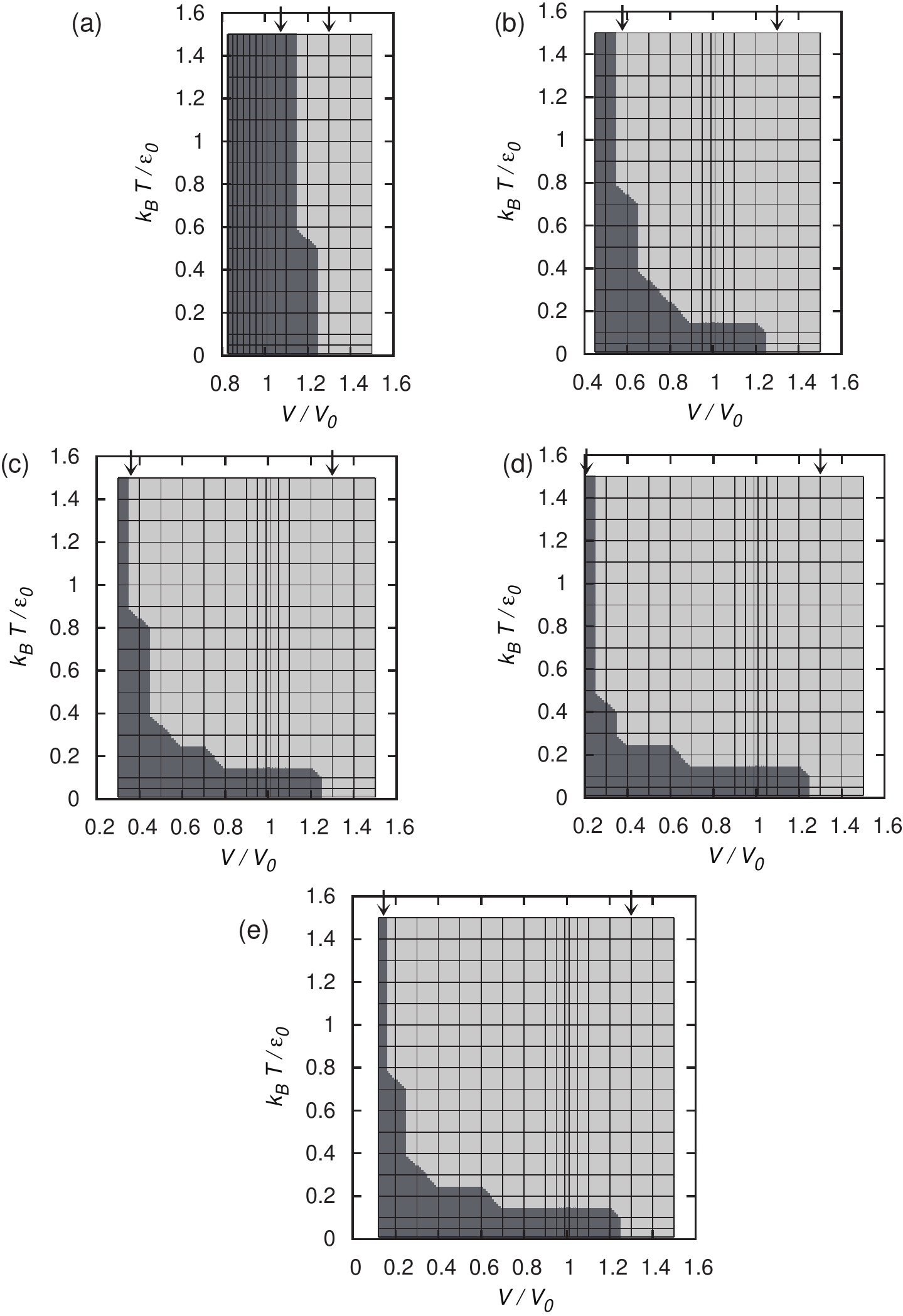}
	\caption{Thermodynamic phase diagrams at $N=1200$. $\sigma_2 / \sigma_1 =$ (a): 1.1, (b): 1.5, (c): 1.9, (d): 2.5, and (e): 3.0. Corresponding thermodynamic phase diagram at $\sigma_2 / \sigma_1 = 2.0$ is given in Fig.~\ref{fig:Frog2DNVTS20-ThermodynamicPhaseDiagram}. Alder transition points of the inner and the outer cores are pointed by arrows above the panel.}
	\label{fig:Frog2DNVTS11S15S19S25S30-ThermodynamicPhaseDiagram}
\end{figure*}

The occurrence probability of percolated clusters and thermodynamic phase diagrams at $\sigma_2 / \sigma_1 = 1.1$, 1.5, 1.9, 2.5, and 3.0 are presented in Figs.~\ref{fig:Frog2DNVTS11S15S19S25S30-PercolationTotal} and \ref{fig:Frog2DNVTS11S15S19S25S30-ThermodynamicPhaseDiagram}. These are qualitatively consistent with the results at $\sigma_2 / \sigma_1 = 2.0$. The Fisher exponent, $\tau = 1.9$, is also kept at these step widths. These results indicate that the discussion on the results at $\sigma_2 / \sigma_1 = 2.0$ also apply to the results at these $\sigma_2 / \sigma_1$.

In addition to the simulation results at $\sigma_2 / \sigma_1 = 2.0$, the simulation results at $\sigma_2 / \sigma_1 = 1.9$ and 2.5 show that the average string length diverges around the region where the melting transition line and the percolation transition line cross~\cite{Norizoe:2005}. A similar behavior is found in Ising spin system where the percolation transition line and the order-disorder line meet at the critical point~\cite{Stauffer1985}.

\section{Conclusions}
\label{sec:Conclusions}
Utilizing Monte Carlo simulation technique, we have studied the phase behavior of 2-dimensional systems of hard particles with square-step repulsive interacting core. The string-like assembly of particles between the two Alder transition densities, each of which corresponds to the inner hard core and the outer hard core, has been found. This string-like assembly is related to the percolation and the critical phenomena. Our model system has recently been studied in experiments, in which the string-like assembly has been confirmed~\cite{Osterman:2007}.

By simulating the system at various step widths, $\sigma_2 / \sigma_1$, we determined dependence of these results on the step width. We have discovered that the string-like assembly, as is shown in Fig.~\ref{fig:Frog2DNVTS20V07T01N10800_005500000MCS}, is observed in the interval $1.7 \lessapprox \sigma_2 / \sigma_1 \lessapprox 2.8$, \textit{i.e.} in the vicinity of $\sigma_2 / \sigma_1 = 2.0$. At this step width $\sigma_2 / \sigma_1 = 2.0$, when the particles are closely and linearly packed on a straight line, both the inner cores and the outer cores of the particles simultaneously pinch and confine each other. Therefore, the region of the string-like assembly ranges around $\sigma_2 / \sigma_1 = 2.0$.

The Fisher exponent of the system is found to be $\tau = 1.9$ at any step width $\sigma_2 / \sigma_1$.

Real systems corresponding to the model particles, which are interacting via the square-step repulsive potential, $\phi (r)$, or the square-step potential with the attractive square-well, were absent in previous studies~\cite{Alder1977,Alder1979,DoubleStepPotential,Malescio:2003,Malescio:2004}. However, in the present study, the interaction potential $\phi (r)$ has been linked to polymer-grafted colloidal particles, \textit{i.e.} real particles, and been derived from the numerical self-consistent field (SCF) calculation~\cite{Norizoe:2005}.

With the use of the SCF method, we are able to calculate the interaction potential between a pair of polymer-grafted colloidal particles for various physical and chemical parameter sets, \textit{e.g.} polymer chain length, block-ratio, grafting density, monomer species, and solvent quality. These results are fed into the particle Monte Carlo scheme. Thus, combining our particle Monte Carlo and SCF calculation results, we can predict and control the phase behaviour of any diblock copolymer-grafted colloidal systems. This methodology allows us to extend this approach to more sophisticated block copolymers and colloidal systems.

The string-like assembly itself is also useful for application. As an example, when the hydrophilic blocks of the grafted polymers that cover the colloid with a thin swollen shell are composed of dielectric material and the thick hydrophobic brush that fills the inside of the shell consists of conductive material, our system can serve as a nano-switching device; the system can conduct the electricity only in the percolated phase. Another example is display devices with use of the difference between the string-like assembly and the disordered phase.

The string-like assembly is also found in 3-dimensional systems~\cite{Norizoe:2005}. A full report will be published in our forthcoming paper~\cite{NorizoeForthcomingPaperFrogNVT3D:2011}.

\begin{acknowledgments}
The authors wish to thank Professors Komajiro Niizeki and Andrei V. Zvelindovsky, who gave us helpful suggestions and discussions.
This work is partially supported by a grant-in-aid for science from the Ministry of Education, Culture, Sports, Science, and Technology, Japan.
\end{acknowledgments}


\begin{thebibliography}{35}%
\makeatletter
\providecommand \@ifxundefined [1]{%
 \@ifx{#1\undefined}
}%
\providecommand \@ifnum [1]{%
 \ifnum #1\expandafter \@firstoftwo
 \else \expandafter \@secondoftwo
 \fi
}%
\providecommand \@ifx [1]{%
 \ifx #1\expandafter \@firstoftwo
 \else \expandafter \@secondoftwo
 \fi
}%
\providecommand \natexlab [1]{#1}%
\providecommand \enquote  [1]{``#1''}%
\providecommand \bibnamefont  [1]{#1}%
\providecommand \bibfnamefont [1]{#1}%
\providecommand \citenamefont [1]{#1}%
\providecommand \href@noop [0]{\@secondoftwo}%
\providecommand \href [0]{\begingroup \@sanitize@url \@href}%
\providecommand \@href[1]{\@@startlink{#1}\@@href}%
\providecommand \@@href[1]{\endgroup#1\@@endlink}%
\providecommand \@sanitize@url [0]{\catcode `\\12\catcode `\$12\catcode
  `\&12\catcode `\#12\catcode `\^12\catcode `\_12\catcode `\%12\relax}%
\providecommand \@@startlink[1]{}%
\providecommand \@@endlink[0]{}%
\providecommand \url  [0]{\begingroup\@sanitize@url \@url }%
\providecommand \@url [1]{\endgroup\@href {#1}{\urlprefix }}%
\providecommand \urlprefix  [0]{URL }%
\providecommand \Eprint [0]{\href }%
\providecommand \doibase [0]{http://dx.doi.org/}%
\providecommand \selectlanguage [0]{\@gobble}%
\providecommand \bibinfo  [0]{\@secondoftwo}%
\providecommand \bibfield  [0]{\@secondoftwo}%
\providecommand \translation [1]{[#1]}%
\providecommand \BibitemOpen [0]{}%
\providecommand \bibitemStop [0]{}%
\providecommand \bibitemNoStop [0]{.\EOS\space}%
\providecommand \EOS [0]{\spacefactor3000\relax}%
\providecommand \BibitemShut  [1]{\csname bibitem#1\endcsname}%
\let\auto@bib@innerbib\@empty
\bibitem [{\citenamefont {Pusey}\ and\ \citenamefont {van
  Megen}(1986)}]{Pusey}%
  \BibitemOpen
  \bibfield  {author} {\bibinfo {author} {\bibfnamefont {P.~N.}\ \bibnamefont
  {Pusey}}\ and\ \bibinfo {author} {\bibfnamefont {W.}~\bibnamefont {van
  Megen}},\ }\href@noop {} {\bibfield  {journal} {\bibinfo  {journal} {Nature}\
  }\textbf {\bibinfo {volume} {320}},\ \bibinfo {pages} {340} (\bibinfo {year}
  {1986})}\BibitemShut {NoStop}%
\bibitem [{\citenamefont {Alder}\ and\ \citenamefont
  {Wainwright}(1960)}]{Alder1960}%
  \BibitemOpen
  \bibfield  {author} {\bibinfo {author} {\bibfnamefont {B.~J.}\ \bibnamefont
  {Alder}}\ and\ \bibinfo {author} {\bibfnamefont {T.~E.}\ \bibnamefont
  {Wainwright}},\ }\href@noop {} {\bibfield  {journal} {\bibinfo  {journal} {J.
  Chem. Phys.}\ }\textbf {\bibinfo {volume} {33}},\ \bibinfo {pages} {1439}
  (\bibinfo {year} {1960})}\BibitemShut {NoStop}%
\bibitem [{\citenamefont {Alder}\ and\ \citenamefont
  {Wainwright}(1962)}]{Alder1962}%
  \BibitemOpen
  \bibfield  {author} {\bibinfo {author} {\bibfnamefont {B.~J.}\ \bibnamefont
  {Alder}}\ and\ \bibinfo {author} {\bibfnamefont {T.~E.}\ \bibnamefont
  {Wainwright}},\ }\href@noop {} {\bibfield  {journal} {\bibinfo  {journal}
  {Phys. Rev.}\ }\textbf {\bibinfo {volume} {127}},\ \bibinfo {pages} {359}
  (\bibinfo {year} {1962})}\BibitemShut {NoStop}%
\bibitem [{\citenamefont {Shull}(1991)}]{Shull:1991}%
  \BibitemOpen
  \bibfield  {author} {\bibinfo {author} {\bibfnamefont {K.~R.}\ \bibnamefont
  {Shull}},\ }\href@noop {} {\bibfield  {journal} {\bibinfo  {journal} {J.
  Chem. Phys.}\ }\textbf {\bibinfo {volume} {94}},\ \bibinfo {pages} {5723}
  (\bibinfo {year} {1991})}\BibitemShut {NoStop}%
\bibitem [{\citenamefont {Matsen}\ and\ \citenamefont
  {Gardiner}(2001)}]{MatsenANDGardiner:2001}%
  \BibitemOpen
  \bibfield  {author} {\bibinfo {author} {\bibfnamefont {M.~W.}\ \bibnamefont
  {Matsen}}\ and\ \bibinfo {author} {\bibfnamefont {J.~M.}\ \bibnamefont
  {Gardiner}},\ }\href@noop {} {\bibfield  {journal} {\bibinfo  {journal} {J.
  Chem. Phys.}\ }\textbf {\bibinfo {volume} {115}},\ \bibinfo {pages} {2794}
  (\bibinfo {year} {2001})}\BibitemShut {NoStop}%
\bibitem [{\citenamefont {Roan}\ and\ \citenamefont
  {Kawakatsu}(2002)}]{RoanANDKawakatsu:2002-1}%
  \BibitemOpen
  \bibfield  {author} {\bibinfo {author} {\bibfnamefont {J.-R.}\ \bibnamefont
  {Roan}}\ and\ \bibinfo {author} {\bibfnamefont {T.}~\bibnamefont
  {Kawakatsu}},\ }\href@noop {} {\bibfield  {journal} {\bibinfo  {journal} {J.
  Chem. Phys.}\ }\textbf {\bibinfo {volume} {116}},\ \bibinfo {pages} {7283}
  (\bibinfo {year} {2002})}\BibitemShut {NoStop}%
\bibitem [{\citenamefont {Borukhov}\ and\ \citenamefont
  {Leibler}(2002)}]{BorukhovANDLeibler:2002}%
  \BibitemOpen
  \bibfield  {author} {\bibinfo {author} {\bibfnamefont {I.}~\bibnamefont
  {Borukhov}}\ and\ \bibinfo {author} {\bibfnamefont {L.}~\bibnamefont
  {Leibler}},\ }\href@noop {} {\bibfield  {journal} {\bibinfo  {journal}
  {Macromolecules}\ }\textbf {\bibinfo {volume} {35}},\ \bibinfo {pages} {5171}
  (\bibinfo {year} {2002})}\BibitemShut {NoStop}%
\bibitem [{\citenamefont {Akcora}\ \emph {et~al.}(2009)\citenamefont {Akcora},
  \citenamefont {Liu}, \citenamefont {Kumar}, \citenamefont {Moll},
  \citenamefont {Li}, \citenamefont {Benicewicz}, \citenamefont {Schadler},
  \citenamefont {Acehan}, \citenamefont {Panagiotopoulos}, \citenamefont
  {Pryamitsyn}, \citenamefont {Ganesan}, \citenamefont {Ilavsky}, \citenamefont
  {Thiyagarajan}, \citenamefont {Colby},\ and\ \citenamefont
  {Douglas}}]{Akcora:2009}%
  \BibitemOpen
  \bibfield  {author} {\bibinfo {author} {\bibfnamefont {P.}~\bibnamefont
  {Akcora}}, \bibinfo {author} {\bibfnamefont {H.}~\bibnamefont {Liu}},
  \bibinfo {author} {\bibfnamefont {S.~K.}\ \bibnamefont {Kumar}}, \bibinfo
  {author} {\bibfnamefont {J.}~\bibnamefont {Moll}}, \bibinfo {author}
  {\bibfnamefont {Y.}~\bibnamefont {Li}}, \bibinfo {author} {\bibfnamefont
  {B.~C.}\ \bibnamefont {Benicewicz}}, \bibinfo {author} {\bibfnamefont
  {L.~S.}\ \bibnamefont {Schadler}}, \bibinfo {author} {\bibfnamefont
  {D.}~\bibnamefont {Acehan}}, \bibinfo {author} {\bibfnamefont {A.~Z.}\
  \bibnamefont {Panagiotopoulos}}, \bibinfo {author} {\bibfnamefont
  {V.}~\bibnamefont {Pryamitsyn}}, \bibinfo {author} {\bibfnamefont
  {V.}~\bibnamefont {Ganesan}}, \bibinfo {author} {\bibfnamefont
  {J.}~\bibnamefont {Ilavsky}}, \bibinfo {author} {\bibfnamefont
  {P.}~\bibnamefont {Thiyagarajan}}, \bibinfo {author} {\bibfnamefont {R.~H.}\
  \bibnamefont {Colby}}, \ and\ \bibinfo {author} {\bibfnamefont {J.~F.}\
  \bibnamefont {Douglas}},\ }\href {\doibase 10.1038/nmat2404} {\bibfield
  {journal} {\bibinfo  {journal} {Nat. Mater.}\ }\textbf {\bibinfo {volume}
  {8}},\ \bibinfo {pages} {354} (\bibinfo {year} {2009})}\BibitemShut {NoStop}%
\bibitem [{\citenamefont {Norizoe}\ and\ \citenamefont
  {Kawakatsu}(2005)}]{Norizoe:2005}%
  \BibitemOpen
  \bibfield  {author} {\bibinfo {author} {\bibfnamefont {Y.}~\bibnamefont
  {Norizoe}}\ and\ \bibinfo {author} {\bibfnamefont {T.}~\bibnamefont
  {Kawakatsu}},\ }\href {\doibase {10.1209/epl/i2005-10288-6}} {\bibfield
  {journal} {\bibinfo  {journal} {{Europhys. Lett.}}\ }\textbf {\bibinfo
  {volume} {72}},\ \bibinfo {pages} {583} (\bibinfo {year} {2005})}\BibitemShut
  {NoStop}%
\bibitem [{\citenamefont {Fleer}, \citenamefont {Stuart},\ and\ \citenamefont
  {Scheutjens}(1993)}]{FleerAndOthers:1993}%
  \BibitemOpen
  \bibfield  {author} {\bibinfo {author} {\bibfnamefont {G.~J.}\ \bibnamefont
  {Fleer}}, \bibinfo {author} {\bibfnamefont {M.~A.~C.}\ \bibnamefont
  {Stuart}}, \ and\ \bibinfo {author} {\bibfnamefont {J.~M. H.~M.}\
  \bibnamefont {Scheutjens}},\ }\href@noop {} {\emph {\bibinfo {title}
  {Polymers at Interfaces}}}\ (\bibinfo  {publisher} {Chapman \& Hall},\
  \bibinfo {year} {1993})\BibitemShut {NoStop}%
\bibitem [{\citenamefont {Israelachvili}(1992)}]{Israelachvili}%
  \BibitemOpen
  \bibfield  {author} {\bibinfo {author} {\bibfnamefont {J.~N.}\ \bibnamefont
  {Israelachvili}},\ }\href@noop {} {\emph {\bibinfo {title} {Intermolecular
  and Surface Forces}}}\ (\bibinfo  {publisher} {Academic Press},\ \bibinfo
  {year} {1992})\BibitemShut {NoStop}%
\bibitem [{\citenamefont {Young}\ and\ \citenamefont
  {Alder}(1977)}]{Alder1977}%
  \BibitemOpen
  \bibfield  {author} {\bibinfo {author} {\bibfnamefont {D.~A.}\ \bibnamefont
  {Young}}\ and\ \bibinfo {author} {\bibfnamefont {B.~J.}\ \bibnamefont
  {Alder}},\ }\href@noop {} {\bibfield  {journal} {\bibinfo  {journal} {Phys.
  Rev. Lett.}\ }\textbf {\bibinfo {volume} {38}},\ \bibinfo {pages} {1213}
  (\bibinfo {year} {1977})}\BibitemShut {NoStop}%
\bibitem [{\citenamefont {Young}\ and\ \citenamefont
  {Alder}(1979)}]{Alder1979}%
  \BibitemOpen
  \bibfield  {author} {\bibinfo {author} {\bibfnamefont {D.~A.}\ \bibnamefont
  {Young}}\ and\ \bibinfo {author} {\bibfnamefont {B.~J.}\ \bibnamefont
  {Alder}},\ }\href@noop {} {\bibfield  {journal} {\bibinfo  {journal} {J.
  Chem. Phys.}\ }\textbf {\bibinfo {volume} {70}},\ \bibinfo {pages} {473}
  (\bibinfo {year} {1979})}\BibitemShut {NoStop}%
\bibitem [{\citenamefont {Buldyrev}\ \emph {et~al.}(2002)\citenamefont
  {Buldyrev}, \citenamefont {Franzese}, \citenamefont {Giovambattista},
  \citenamefont {Malescio}, \citenamefont {Sadr-Lahijany}, \citenamefont
  {Scala}, \citenamefont {Skibinsky},\ and\ \citenamefont
  {Stanley}}]{DoubleStepPotential}%
  \BibitemOpen
  \bibfield  {author} {\bibinfo {author} {\bibfnamefont {S.}~\bibnamefont
  {Buldyrev}}, \bibinfo {author} {\bibfnamefont {G.}~\bibnamefont {Franzese}},
  \bibinfo {author} {\bibfnamefont {N.}~\bibnamefont {Giovambattista}},
  \bibinfo {author} {\bibfnamefont {G.}~\bibnamefont {Malescio}}, \bibinfo
  {author} {\bibfnamefont {M.}~\bibnamefont {Sadr-Lahijany}}, \bibinfo {author}
  {\bibfnamefont {A.}~\bibnamefont {Scala}}, \bibinfo {author} {\bibfnamefont
  {A.}~\bibnamefont {Skibinsky}}, \ and\ \bibinfo {author} {\bibfnamefont
  {H.}~\bibnamefont {Stanley}},\ }\href@noop {} {\bibfield  {journal} {\bibinfo
   {journal} {Phys. A}\ }\textbf {\bibinfo {volume} {304}},\ \bibinfo {pages}
  {23} (\bibinfo {year} {2002})}\BibitemShut {NoStop}%
\bibitem [{\citenamefont {Jayaraman}, \citenamefont {Newton},\ and\
  \citenamefont {McDonough}(1967)}]{Cesium1967}%
  \BibitemOpen
  \bibfield  {author} {\bibinfo {author} {\bibfnamefont {A.}~\bibnamefont
  {Jayaraman}}, \bibinfo {author} {\bibfnamefont {R.~C.}\ \bibnamefont
  {Newton}}, \ and\ \bibinfo {author} {\bibfnamefont {J.~M.}\ \bibnamefont
  {McDonough}},\ }\href@noop {} {\bibfield  {journal} {\bibinfo  {journal}
  {Phys. Rev.}\ }\textbf {\bibinfo {volume} {159}},\ \bibinfo {pages} {527}
  (\bibinfo {year} {1967})}\BibitemShut {NoStop}%
\bibitem [{\citenamefont {King}\ \emph {et~al.}(1970)\citenamefont {King},
  \citenamefont {Lee}, \citenamefont {Harris},\ and\ \citenamefont
  {Smith}}]{Cerium1970}%
  \BibitemOpen
  \bibfield  {author} {\bibinfo {author} {\bibfnamefont {E.}~\bibnamefont
  {King}}, \bibinfo {author} {\bibfnamefont {J.~A.}\ \bibnamefont {Lee}},
  \bibinfo {author} {\bibfnamefont {I.~R.}\ \bibnamefont {Harris}}, \ and\
  \bibinfo {author} {\bibfnamefont {T.~F.}\ \bibnamefont {Smith}},\ }\href@noop
  {} {\bibfield  {journal} {\bibinfo  {journal} {Phys. Rev. B}\ }\textbf
  {\bibinfo {volume} {1}},\ \bibinfo {pages} {1380} (\bibinfo {year}
  {1970})}\BibitemShut {NoStop}%
\bibitem [{\citenamefont {Malescio}\ and\ \citenamefont
  {Pellicane}(2003)}]{Malescio:2003}%
  \BibitemOpen
  \bibfield  {author} {\bibinfo {author} {\bibfnamefont {G.}~\bibnamefont
  {Malescio}}\ and\ \bibinfo {author} {\bibfnamefont {G.}~\bibnamefont
  {Pellicane}},\ }\href@noop {} {\bibfield  {journal} {\bibinfo  {journal}
  {Nat. Mater.}\ }\textbf {\bibinfo {volume} {2}},\ \bibinfo {pages} {97}
  (\bibinfo {year} {2003})}\BibitemShut {NoStop}%
\bibitem [{\citenamefont {Malescio}\ and\ \citenamefont
  {Pellicane}(2004)}]{Malescio:2004}%
  \BibitemOpen
  \bibfield  {author} {\bibinfo {author} {\bibfnamefont {G.}~\bibnamefont
  {Malescio}}\ and\ \bibinfo {author} {\bibfnamefont {G.}~\bibnamefont
  {Pellicane}},\ }\href@noop {} {\bibfield  {journal} {\bibinfo  {journal}
  {Phys. Rev. E}\ }\textbf {\bibinfo {volume} {70}},\ \bibinfo {pages} {021202}
  (\bibinfo {year} {2004})}\BibitemShut {NoStop}%
\bibitem [{\citenamefont {Norizoe}\ and\ \citenamefont
  {Kawakatsu}(2003{\natexlab{a}})}]{Norizoe:2003April}%
  \BibitemOpen
  \bibfield  {author} {\bibinfo {author} {\bibfnamefont {Y.}~\bibnamefont
  {Norizoe}}\ and\ \bibinfo {author} {\bibfnamefont {T.}~\bibnamefont
  {Kawakatsu}},\ }\href@noop {} {\enquote {\bibinfo {title} {Monte-carlo
  simulation of string-like colloidal assembly},}\ }\bibinfo {howpublished}
  {Presentation at 298. WE-Heraeus Seminar ``New Approaches and Perspectives in
  Polymer Physics''},\ \bibinfo {address} {Physikzentrum Bad Honnef (Germany)}
  (\bibinfo {year} {2003}{\natexlab{a}})\BibitemShut {NoStop}%
\bibitem [{\citenamefont {Norizoe}\ and\ \citenamefont
  {Kawakatsu}(2003{\natexlab{b}})}]{Norizoe:2003November}%
  \BibitemOpen
  \bibfield  {author} {\bibinfo {author} {\bibfnamefont {Y.}~\bibnamefont
  {Norizoe}}\ and\ \bibinfo {author} {\bibfnamefont {T.}~\bibnamefont
  {Kawakatsu}},\ }\href@noop {} {\enquote {\bibinfo {title} {Monte-carlo
  simulation of string-like colloidal assembly},}\ }\bibinfo {howpublished}
  {Presentation at The 3rd International Symposium on Slow Dynamics in Complex
  Systems},\ \bibinfo {address} {Tohoku University, Sendai (Japan)} (\bibinfo
  {year} {2003}{\natexlab{b}})\BibitemShut {NoStop}%
\bibitem [{\citenamefont {Norizoe}(2004)}]{Master'sThesis}%
  \BibitemOpen
  \bibfield  {author} {\bibinfo {author} {\bibfnamefont {Y.}~\bibnamefont
  {Norizoe}},\ }\emph {\bibinfo {title} {Monte-{C}arlo Simulation of
  String-like Colloidal Assembly}},\ \href@noop {} {Master's thesis},\ \bibinfo
   {school} {Department of Physics, Tohoku University}, \bibinfo {address}
  {Sendai, Japan} (\bibinfo {year} {2004})\BibitemShut {NoStop}%
\bibitem [{\citenamefont {Glaser}\ \emph {et~al.}(2007)\citenamefont {Glaser},
  \citenamefont {Grason}, \citenamefont {Kamien}, \citenamefont
  {Ko{\v{s}}mrlj}, \citenamefont {Santangelo},\ and\ \citenamefont
  {Ziherl}}]{Glaser:2007}%
  \BibitemOpen
  \bibfield  {author} {\bibinfo {author} {\bibfnamefont {M.~A.}\ \bibnamefont
  {Glaser}}, \bibinfo {author} {\bibfnamefont {G.~M.}\ \bibnamefont {Grason}},
  \bibinfo {author} {\bibfnamefont {R.~D.}\ \bibnamefont {Kamien}}, \bibinfo
  {author} {\bibfnamefont {A.}~\bibnamefont {Ko{\v{s}}mrlj}}, \bibinfo {author}
  {\bibfnamefont {C.~D.}\ \bibnamefont {Santangelo}}, \ and\ \bibinfo {author}
  {\bibfnamefont {P.}~\bibnamefont {Ziherl}},\ }\href@noop {} {\bibfield
  {journal} {\bibinfo  {journal} {EPL}\ }\textbf {\bibinfo {volume} {78}},\
  \bibinfo {pages} {46004} (\bibinfo {year} {2007})}\BibitemShut {NoStop}%
\bibitem [{\citenamefont {Fomin}\ \emph {et~al.}(2008)\citenamefont {Fomin},
  \citenamefont {Gribova}, \citenamefont {Ryzhov}, \citenamefont {Stishov},\
  and\ \citenamefont {Frenkel}}]{Fomin:2008}%
  \BibitemOpen
  \bibfield  {author} {\bibinfo {author} {\bibfnamefont {Y.~D.}\ \bibnamefont
  {Fomin}}, \bibinfo {author} {\bibfnamefont {N.~V.}\ \bibnamefont {Gribova}},
  \bibinfo {author} {\bibfnamefont {V.~N.}\ \bibnamefont {Ryzhov}}, \bibinfo
  {author} {\bibfnamefont {S.~M.}\ \bibnamefont {Stishov}}, \ and\ \bibinfo
  {author} {\bibfnamefont {D.}~\bibnamefont {Frenkel}},\ }\href {\doibase
  10.1063/1.2965880} {\bibfield  {journal} {\bibinfo  {journal} {J. Chem.
  Phys.}\ }\textbf {\bibinfo {volume} {129}},\ \bibinfo {eid} {064512}
  (\bibinfo {year} {2008})}\BibitemShut {NoStop}%
\bibitem [{\citenamefont {Camp}(2003)}]{Camp:2003}%
  \BibitemOpen
  \bibfield  {author} {\bibinfo {author} {\bibfnamefont {P.~J.}\ \bibnamefont
  {Camp}},\ }\href@noop {} {\bibfield  {journal} {\bibinfo  {journal} {Phys.
  Rev. E}\ }\textbf {\bibinfo {volume} {68}},\ \bibinfo {pages} {061506}
  (\bibinfo {year} {2003})}\BibitemShut {NoStop}%
\bibitem [{\citenamefont {Pauschenwein}\ and\ \citenamefont
  {Kahl}(2008)}]{Pauschenwein:2008}%
  \BibitemOpen
  \bibfield  {author} {\bibinfo {author} {\bibfnamefont {G.~J.}\ \bibnamefont
  {Pauschenwein}}\ and\ \bibinfo {author} {\bibfnamefont {G.}~\bibnamefont
  {Kahl}},\ }\href {\doibase 10.1063/1.3006065} {\bibfield  {journal} {\bibinfo
   {journal} {J. Chem. Phys.}\ }\textbf {\bibinfo {volume} {129}},\ \bibinfo
  {eid} {174107} (\bibinfo {year} {2008})}\BibitemShut {NoStop}%
\bibitem [{\citenamefont {Norizoe}\ and\ \citenamefont
  {Kawakatsu}()}]{NorizoeForthcomingPaperFrogNVT3D:2011}%
  \BibitemOpen
  \bibfield  {author} {\bibinfo {author} {\bibfnamefont {Y.}~\bibnamefont
  {Norizoe}}\ and\ \bibinfo {author} {\bibfnamefont {T.}~\bibnamefont
  {Kawakatsu}},\ }\href@noop {} {\enquote {\bibinfo {title} {Particle {M}onte
  {C}arlo simulation of string-like colloidal assembly in 3 dimensions},}\
  }\bibinfo {note} {Unpublished}\BibitemShut {NoStop}%
\bibitem [{\citenamefont {Allen}\ and\ \citenamefont
  {Tildesley}(1989)}]{ComputerSimulationOfLiquids}%
  \BibitemOpen
  \bibfield  {author} {\bibinfo {author} {\bibfnamefont {M.~P.}\ \bibnamefont
  {Allen}}\ and\ \bibinfo {author} {\bibfnamefont {D.~J.}\ \bibnamefont
  {Tildesley}},\ }\href@noop {} {\emph {\bibinfo {title} {Computer Simulation
  of Liquids}}}\ (\bibinfo  {publisher} {Oxford University Press},\ \bibinfo
  {address} {Oxford},\ \bibinfo {year} {1989})\BibitemShut {NoStop}%
\bibitem [{\citenamefont {Frenkel}\ and\ \citenamefont
  {Smit}(2002)}]{Frenkel:UnderstandingMolecularSimulation2002}%
  \BibitemOpen
  \bibfield  {author} {\bibinfo {author} {\bibfnamefont {D.}~\bibnamefont
  {Frenkel}}\ and\ \bibinfo {author} {\bibfnamefont {B.}~\bibnamefont {Smit}},\
  }\href@noop {} {\emph {\bibinfo {title} {Understanding molecular simulation:
  from algorithms to applications}}}\ (\bibinfo  {publisher} {Academic Press},\
  \bibinfo {address} {London},\ \bibinfo {year} {2002})\BibitemShut {NoStop}%
\bibitem [{\citenamefont {Matsumoto}\ and\ \citenamefont
  {Nishimura}(1998)}]{MersenneTwister1}%
  \BibitemOpen
  \bibfield  {author} {\bibinfo {author} {\bibfnamefont {M.}~\bibnamefont
  {Matsumoto}}\ and\ \bibinfo {author} {\bibfnamefont {T.}~\bibnamefont
  {Nishimura}},\ }\href {\doibase http://doi.acm.org/10.1145/272991.272995}
  {\bibfield  {journal} {\bibinfo  {journal} {ACM Trans. Model. Comput.
  Simul.}\ }\textbf {\bibinfo {volume} {8}},\ \bibinfo {pages} {3} (\bibinfo
  {year} {1998})}\BibitemShut {NoStop}%
\bibitem [{\citenamefont {Matsumoto}\ and\ \citenamefont
  {Kurita}(1992)}]{MersenneTwister2}%
  \BibitemOpen
  \bibfield  {author} {\bibinfo {author} {\bibfnamefont {M.}~\bibnamefont
  {Matsumoto}}\ and\ \bibinfo {author} {\bibfnamefont {Y.}~\bibnamefont
  {Kurita}},\ }\href {\doibase http://doi.acm.org/10.1145/146382.146383}
  {\bibfield  {journal} {\bibinfo  {journal} {ACM Trans. Model. Comput.
  Simul.}\ }\textbf {\bibinfo {volume} {2}},\ \bibinfo {pages} {179} (\bibinfo
  {year} {1992})}\BibitemShut {NoStop}%
\bibitem [{\citenamefont {Matsumoto}\ and\ \citenamefont
  {Kurita}(1994)}]{MersenneTwister3}%
  \BibitemOpen
  \bibfield  {author} {\bibinfo {author} {\bibfnamefont {M.}~\bibnamefont
  {Matsumoto}}\ and\ \bibinfo {author} {\bibfnamefont {Y.}~\bibnamefont
  {Kurita}},\ }\href {\doibase http://doi.acm.org/10.1145/189443.189445}
  {\bibfield  {journal} {\bibinfo  {journal} {ACM Trans. Model. Comput.
  Simul.}\ }\textbf {\bibinfo {volume} {4}},\ \bibinfo {pages} {254} (\bibinfo
  {year} {1994})}\BibitemShut {NoStop}%
\bibitem [{\citenamefont {Osterman}\ \emph {et~al.}(2007)\citenamefont
  {Osterman}, \citenamefont {Babi{\v{c}}}, \citenamefont {Poberaj},
  \citenamefont {Dobnikar},\ and\ \citenamefont {Ziherl}}]{Osterman:2007}%
  \BibitemOpen
  \bibfield  {author} {\bibinfo {author} {\bibfnamefont {N.}~\bibnamefont
  {Osterman}}, \bibinfo {author} {\bibfnamefont {D.}~\bibnamefont
  {Babi{\v{c}}}}, \bibinfo {author} {\bibfnamefont {I.}~\bibnamefont
  {Poberaj}}, \bibinfo {author} {\bibfnamefont {J.}~\bibnamefont {Dobnikar}}, \
  and\ \bibinfo {author} {\bibfnamefont {P.}~\bibnamefont {Ziherl}},\ }\href
  {\doibase 10.1103/PhysRevLett.99.248301} {\bibfield  {journal} {\bibinfo
  {journal} {Phys. Rev. Lett.}\ }\textbf {\bibinfo {volume} {99}},\ \bibinfo
  {pages} {248301} (\bibinfo {year} {2007})}\BibitemShut {NoStop}%
\bibitem [{\citenamefont {{Norizoe}}\ and\ \citenamefont
  {{Kawakatsu}}(2010)}]{2010:NorizoeMuPTLetterArXiv}%
  \BibitemOpen
  \bibfield  {author} {\bibinfo {author} {\bibfnamefont {Y.}~\bibnamefont
  {{Norizoe}}}\ and\ \bibinfo {author} {\bibfnamefont {T.}~\bibnamefont
  {{Kawakatsu}}},\ }\href@noop {} {\bibfield  {journal} {\bibinfo  {journal}
  {ArXiv e-prints}\ } (\bibinfo {year} {2010})},\ \Eprint
  {http://arxiv.org/abs/1011.3205} {arXiv:1011.3205 [cond-mat.soft]}
  \BibitemShut {NoStop}%
\bibitem [{\citenamefont {{Norizoe}}\ and\ \citenamefont
  {{Kawakatsu}}(2011)}]{2011:NorizoeMuPTFullArXiv}%
  \BibitemOpen
  \bibfield  {author} {\bibinfo {author} {\bibfnamefont {Y.}~\bibnamefont
  {{Norizoe}}}\ and\ \bibinfo {author} {\bibfnamefont {T.}~\bibnamefont
  {{Kawakatsu}}},\ }\href@noop {} {\bibfield  {journal} {\bibinfo  {journal}
  {ArXiv e-prints}\ } (\bibinfo {year} {2011})},\ \Eprint
  {http://arxiv.org/abs/1101.1695} {arXiv:1101.1695 [physics.comp-ph]}
  \BibitemShut {NoStop}%
\bibitem [{\citenamefont {Stauffer}\ and\ \citenamefont
  {Aharony}(1994)}]{Stauffer1985}%
  \BibitemOpen
  \bibfield  {author} {\bibinfo {author} {\bibfnamefont {D.}~\bibnamefont
  {Stauffer}}\ and\ \bibinfo {author} {\bibfnamefont {A.}~\bibnamefont
  {Aharony}},\ }\href@noop {} {\emph {\bibinfo {title} {Introduction to
  Percolation Theory}}}\ (\bibinfo  {publisher} {Taylor \& Francis},\ \bibinfo
  {address} {London},\ \bibinfo {year} {1994})\BibitemShut {NoStop}%
\end{thebibliography}
%

\end{document}